\documentclass[conference,a4paper]{IEEEtran}
\IEEEoverridecommandlockouts

\makeatletter
\newcommand{\newlineauthors}{%
  \end{@IEEEauthorhalign}\hfill\mbox{}\par
  \mbox{}\hfill\begin{@IEEEauthorhalign}
}
\usepackage{cite}
\usepackage{amsmath,amssymb,amsfonts}
\usepackage{algorithmic}
\usepackage{graphicx}
\usepackage{textcomp}
\usepackage{xcolor}
\usepackage[hyphens]{url}
\usepackage{breakurl}
\usepackage{anyfontsize}
\usepackage{multirow,multicol, makecell, booktabs}
\def\BibTeX{{\rm B\kern-.05em{\sc i\kern-.025em b}\kern-.08em
    T\kern-.1667em\lower.7ex\hbox{E}\kern-.125emX}}
\begin{document}

\makeatletter
\def\ps@IEEEtitlepagestyle{%
  \def\@oddfoot{\mycopyrightnotice}%
  \def\@evenfoot{}%
}
\def\mycopyrightnotice{%
  {\footnotesize 978-8-8872-3749-8 \copyright 2020 AEIT \hfill} 
}
\makeatother

\title{Car-Driver Drowsiness Assessment through 1D Temporal Convolutional Networks \\
}

\author{\IEEEauthorblockN{Francesco Rundo}
\IEEEauthorblockA{\normalsize{ADG Central R\&D Division}\\
\textit{STMicroelectronics}\\
Catania, Italy \\
francesco.rundo@st.com}
\and
\IEEEauthorblockN{Concetto Spampinato}
\IEEEauthorblockA{\textit{Perceive Lab}\\
\textit{University of Catania} \\
Catania, Italy \\
cspampinato@dieii.unict.it }
\and
\IEEEauthorblockN{Michael Rundo}
\IEEEauthorblockA{\textit{Perceive Lab}\\
\textit{University of Catania} \\
Catania, Italy \\
michael.rundo3764@gmail.com }

}

\maketitle

\begin{abstract}
Recently, the scientific progress of Advanced Driver Assistance System solutions (ADAS) has played a key role in enhancing the overall safety of driving. ADAS technology enables active control of vehicles to prevent potentially risky situations. An important aspect that researchers have focused on is the analysis of the driver's attention level, as recent reports confirmed a rising number of accidents caused by drowsiness or lack of attentiveness.
To address this issue, various studies have suggested monitoring the driver's physiological state, as there exists a well-established connection between the Autonomic Nervous System (ANS) and the level of attention. For our study, we designed an innovative bio-sensor comprising near-infrared LED emitters and photo-detectors, specifically a Silicon PhotoMultiplier device. This allowed us to assess the driver's physiological status by analyzing the associated PhotoPlethysmography (PPG) signal.Furthermore, we developed an embedded time-domain hyper-filtering technique in conjunction with a 1D Temporal Convolutional architecture that embdes a progressive dilation setup. This integrated system enables near real-time classification of driver drowsiness, yielding remarkable accuracy levels of approximately 96\%.
\end{abstract}

\begin{IEEEkeywords}
Drowsiness, Deep learning, D-CNN, Deep-LSTM, PPG (PhotoPlethySmography)
\end{IEEEkeywords}

\section{Introduction}
In the medical domain, the term "drowsiness" refers to a state characterized by a reduced level of alertness and a tendency to sleep. The advancement of technology has motivated researchers to develop effective methods for detecting critical levels of driver drowsiness, aiming to prevent serious road traffic accidents. Numerous studies have extensively explored the correlation between attention level and Heart Rate Variability (HRV) [1]. HRV serves as an indicator of the autonomic regulation of the heart and can be derived from the frequency analysis of the ElectroCardioGraphy (ECG) or PhotoPlethysmoGraphic (PPG) signals obtained from the subject. Essentially, HRV reflects the variability in the time intervals between consecutive heartbeats, predominantly influenced by the dynamic interplay between the Autonomous Nervous System (ANS) and the heart [1]. Recent scientific investigations have focused on analyzing the relationship between drowsiness and ANS activity through HRV. In this study, we propose an innovative pipeline for evaluating the attention level of car drivers using PPG signals. The organization of this paper is as follows: Section II presents an overview of related works. In Section III, we provide detailed information about the hardware framework employed to acquire the PPG signals, along with the associated processing pipeline. Section IV describes the Deep Learning framework utilized for classifying the collected PPG waveforms. Finally, we present the experimental results and discuss future avenues of research.

\section{Related Works}

Numerous studies have focused on utilizing physiological signals to evaluate the driver's condition. In a study by Vicente et al. \cite{vicente2011detection}, the authors employed Electrocardiography (ECG) signals to classify the driver's drowsiness status. However, ECG signals are susceptible to artifacts that can distort the accurate acquisition of Heart Rate Variability (HRV). To mitigate this issue, the authors proposed a pipeline to filter and stabilize the ECG signal. Many existing approaches concentrate on detecting the attention level through the analysis of Lower-Frequency (LF) and Higher-Frequency (HF) features. In a study by Szypulska et al. \cite{szypulska2012prediction}, the authors developed a drowsiness detection system that assessed fatigue and sleep onset by examining the LF/HF ratio derived from frequency analysis of the ECG's R-R tachogram. The results demonstrated the efficacy of this approach in Advanced Driver Assistance Systems (ADAS) applications. However, these methods require the acquisition of ECG signals for HRV analysis.
Traditional ECG signal acquisition involves the configuration of Einthoven's Triangle \cite{abi2019einthoven}, necessitating the driver's contact with three specific electrodes to ensure accurate signal acquisition. Consequently, this introduces challenges in maintaining the robustness of the ECG signal sampling system, which in turn affects the quality of the derived HRV signal \cite{abi2019einthoven}. In light of these challenges, recent studies have shifted towards the utilization of PhotoPlethysmoGraphic (PPG) signals as an alternative to ECG signals \cite{rundo2018advanced}. Unlike ECG signals, PPG signals only require a single contact point on the subject's skin for accurate sampling \cite{rundo2018advanced}.In recent years, Deep Learning (DL) approaches have gained significant attention due to their effectiveness in estimating a subject's drowsiness. For instance, studies by Hong et al. \cite{hong2007drivers} and Alshaqaqi et al. \cite{alshaqaqi2013driver} propose DL architectures to track changes in eye state for estimating driver drowsiness. However, these approaches face challenges in recording video sequences under adverse conditions, such as varying illumination and occlusions.Driven by the rapid advancements in Machine Learning, researchers have developed effective architectures for classifying the driver's physiological status. Cheon et al. \cite{cheon2017sensor} introduced a pipeline that addresses the classification problem of driver drowsiness by utilizing a variety of sensors on the steering wheel. Additionally, Choi et al. \cite{choi2018driver} employed a Multimodal Deep Learning model to recognize the driver's vigilance level by analyzing both visual and physiological data. While previous methods have shown impressive results, they often involve substantial computational costs, limiting their applicability on resource-constrained automotive-grade devices.

\begin{figure}
\centerline{\includegraphics[ width=0.8\columnwidth]{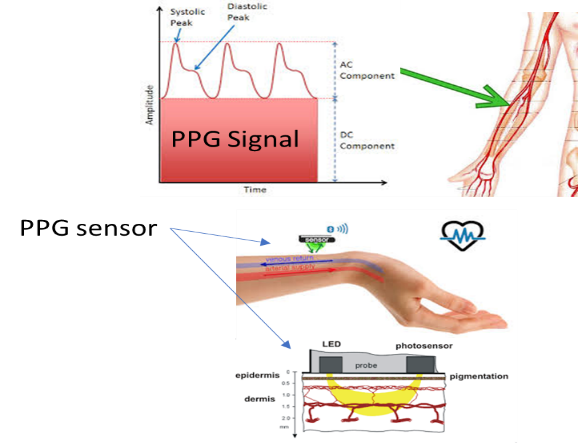}}
\caption{The proposed PPG sampling system.}
\label{fig1}
\end{figure}

\begin{table*}
\caption{Low-Pass and High-Pass Filter Design for the PPG Signal}
\begin{center}
\resizebox{0.8\textwidth}{!}{%
\begin{tabular}{|c|c|c|c|c|}
\hline
\rule{0pt}{3ex}
\large\textbf{Type} & \large\textit{Frequency pass [Hz]} & \large\textit{Frequency stop [Hz]} & \large\textit{Passband Attenuation [dB]} & \large\textit{Stopband Attenuation [dB]} \\ 
\hline
\rule{0pt}{3ex}
\large LP & \large 4.8 & \large 10 & \large 0.001 & \large 100 \\
\hline
\rule{0pt}{3ex}
\large HP & \large 1 & \large 0.3 & \large 0.01 & \large 40 \\
\hline
\end{tabular}}
\label{tab1}
\end{center}
\end{table*}

\section{Methods and Materials}
\begin{figure}
\centerline{\includegraphics[ width=0.9\columnwidth]{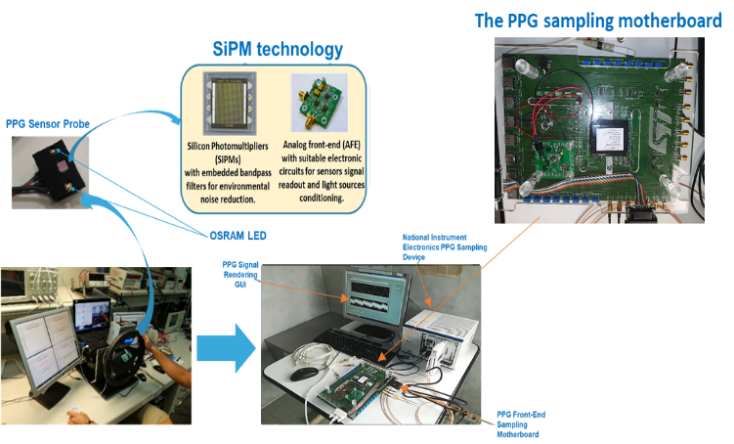}}
\caption{The proposed driver drowsiness monitoring pipeline.}
\label{fig2}
\end{figure}

In this investigation, our focus was on developing a system to capture the PPG signal of car drivers, enabling us to evaluate and monitor their corresponding attention levels. The PPG signal is a non-invasive approach widely used for analyzing the heart's pulse rate. By extracting features from the PPG signal, we can monitor various factors such as heart pulse, respiratory rate, as well as vascular and cardiac disorders \cite{rundo2018advanced}.The PPG waveform consists of two main components: the pulsatile 'AC' signal, which represents the cardiac-synchronous changes in blood volume, and the 'DC' component, which is influenced by the processes of respiration and thermo-regulation. During a cardiac cycle, as the heart pumps blood towards the peripheral regions of the body, it generates pressure that causes the arteries and arterioles in the subcutaneous tissue to expand. This expansion, in turn, results in an increase in blood volume. To capture these changes in volume, we utilize a specialized device equipped with a light-emitting component and a detector that is placed on the skin. By illuminating the skin and measuring the amount of back-scattered light received by the detector \cite{rundo2018advanced}, we can effectively detect the heart's pressure pulse, which is manifested as a peak in the PPG waveform. In Fig.~\ref{fig1}, the process underlying PPG waveform(s) formation is reported. To perform the PPG acquisition, we used the PPG sampling device composed of the Silicon Photomultiplier sensor \cite{vinciguerra2018ppg, mazzillo2018characterization}. The proposed PPG probes comprises an array device, called Silicon Photomultipliers (SiPMs) \cite{conoci2018live}, characterized by a total area of $4.0\times4.5$ $mm^{2}$ and $4871$ square microcells with $60$ µm pitch. The devices present a geometrical fill factor of $67.4\%$ and are packaged in a surface mount housing (SMD) with about $5.1\times5.1$ $mm^{2}$ total area \cite{rundo2018advanced}.For the PPG signal acquisition system, we employed a Pixelteq dichroic bandpass filter with specific characteristics. The filter was designed with a pass-band centered at approximately $540$ nm and a Full Width at Half Maximum (FWHM) of $70$ nm. Its optical transmission in the pass-band range exceeded $90-95\%$. To secure the filter onto the SMD package, we used a Loctite 352TM adhesive. The PPG detector comprises a light emitter and a detector based on advanced technology. We utilized OSRAM LT M673 LEDs, which integrate InGan technology, in a compact SMD package. These LEDs offer a wide-angle view of 120° and cover an area of $2.3\times1.5$ $mm^{2}$. With a spectral bandwidth of 33 nm, they emit lower power in the standard range.To optimize the functionality of the PPG probe, we designed a printed circuit board (PCB) that incorporates a user-interface based on National Instruments (NI) instrumentation. The PCB features a 4V portable battery, power management circuits, a conditioning circuit for the output signals of the SiPMs, as well as multiple USB connectors for PPG probes and corresponding SMA output connectors. In our previous works \cite{rundo2018advanced, conoci2018live}, we provided additional information regarding the hardware setup for PPG signal acquisition. The PPG Sensor Probe consists of the SiPM sensor and the corresponding LEDs. To manage the power consumption of the SiPM device, we implemented a Power Management Circuit \cite{vinciguerra2018ppg, mazzillo2018characterization, conoci2018live}. For PPG signal acquisition, we placed the PPG sensor probe on the car's steering wheel. The driver was instructed to maintain their hand in contact with the probe to trigger the signal. The acquired PPG raw signal was processed internally by the NI device, which includes multiple 24-bit ADCs. Furthermore, the NI device is equipped with a Windows-based operating system and utilizes the LabView software framework \cite{rundo2018advanced}.
Figure~\ref{fig2} illustrates the pipeline employed to process the PPG signal using the NI device. We developed a LabView-based algorithm to filter the raw PPG data, implementing both low-pass and high-pass FIR (Finite Impulse Response) filters. Additionally, we computed the first and second derivatives of the PPG signal as part of the preprocessing step, enabling the evaluation of the minimum and maximum extremes for each waveform. The rendered PPG signal was displayed on the monitor connected to the NI device, as depicted in Figure~\ref{fig2}. More detailed descriptions regarding the NI device and LabView software framework can be found in our previous publications \cite{rundo2018advanced, vinciguerra2018ppg}. For the purpose of this study, we utilized the MATLAB framework to process the PPG raw data, as outlined in the validation session.

\subsection{The Hyper Filtering Layers}
To mitigate the impact of motion and noise artifacts in the acquired PPG signal, we employed a frequency filtering approach and developed a signal stabilization algorithm. In our bio-inspired pipeline, which has shown promising results in our previous works~\cite{rundo2018advanced}, \cite{vinciguerra2018ppg, mazzillo2018characterization, conoci2018live}, we utilized a series of FIR filters to implement a low-pass and high-pass filtering scheme within the frequency range of $1-10$ Hz. This effectively eliminated unwanted noise components, including the interference caused by the 50 Hz power line frequency. To further enhance the robustness of the pipeline, we introduced a set of hyper-filtering layers. The objective was to explore the potential of hyperspectral imaging concepts in the context of PPG signal processing.
Drawing inspiration from Chang's work~\cite{chang2003hyperspectral}, we adapted the concept of hyperspectral imaging to analyze 1D signals, specifically the PPG signal. Hyperspectral imaging is a technique used to capture visual information from the entire electromagnetic spectrum, allowing for detailed spectral analysis of each pixel in an image. While traditionally applied to visual imagery for object recognition and material identification, we explored its application to 1D signals. Our objective was to investigate whether collecting information from multiple frequency ranges of the PPG signal, akin to the concept of hyperspectral imaging, could provide more comprehensive features related to the driver's attention level. Instead of applying a single set of filters (e.g., low-pass and high-pass), we analyzed a range of frequencies within the 1-10 Hz range to characterize the individual PPG waveforms. Through our investigation, we discovered that the informative frequency range for attention assessment fell within this range. To further enhance the hyper-filtering process, we explored the possibility of subdividing the frequency range into sub-intervals. This subdivision aimed to simulate a hyper-spectral process, where each sub-interval captured specific frequency components related to attention. Consequently, we incorporated both low-pass and high-pass filters in our configuration, implementing two layers of hyper-filtering. In the hyper low-pass filtering layer, we adjusted the frequencies in the low-pass part while keeping the cutoff frequency of the high-pass filter constant, and vice versa. To ensure optimal performance and minimize the introduction of noise artifacts, we opted to use Butterworth filters in both layers of hyper-filtering~\cite{rundo2019ad, bianchi2007electronic}.
To determine the optimal number of sub-intervals within the 1-10 Hz frequency range, we employed a "try-and-error" approach combined with heuristic tests. Our goal was to strike a balance between computational load and discriminative capacity, ensuring that the sub-intervals effectively captured the relevant frequency components while avoiding excessive computational complexity.
Through iterative experimentation and analysis, we determined that a total of 11 sub-intervals provided the most favorable results.  Once subdividing the frequency into $11$ sub-bands, we designed a Reinforcement Learning (RL) algorithm. The implementation details of this approach are reported in the following items:
\begin{itemize}
    \item We defined an action $a_{t}$ as the sub-band frequency selected in the range reported in Table I and according to the type of filtering (low-pass or high-pass);
    \item an Agent is defined selecting the action $a_{t}$
    \item We defined a next state $S_{t+1}$ as a set of pre-processed signals obtained collecting the value of each input PPG samples (in a windows of 5 sec sampling at 1 Khz as sampling frequency) of the filtered PPG raw signal at specific sub-band frequency of the action $a_{t}$;
    \item We define an environment Reward as $R(.|s_{t},a_{t})$ i.e., a measure of drowsiness of the car driver. We indicated as $R(.|s_{t},a_{t})$ the distance of the output of the deep learning system (regression layer plus SoftMax classification) with respect car-driver’s level of attention.
\end{itemize}
We determined the optimal policy $P_{o}$ that minimizes the cumulative discount reward by applying the following formula:
\begin{equation}
    P_o=argmax_{P_o}\ E\left[\sum_{t\geq0}{\gamma^tR\left(.\middle|s_t,a_t\right)|P_o\ }\right]
\end{equation}

\begin{figure}
\centerline{\includegraphics[ width=\columnwidth]{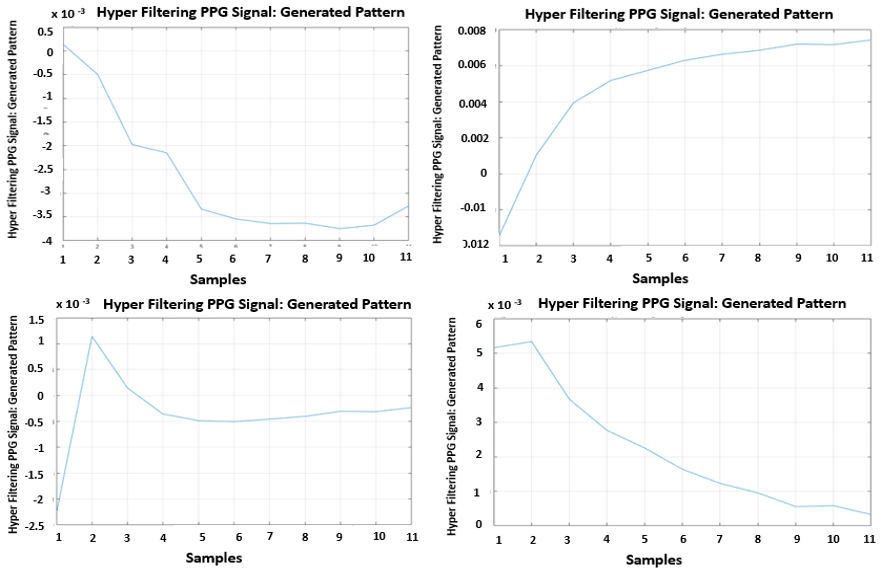}}
\caption{Some instances of the hyper-filtered PPG generated patterns}
\label{fig5}
\end{figure}

Where ${\gamma}$ is a proper discounted coefficient in (0,1). In order to evaluate the the goodness of a state st and the goodness of a state-action couple $(s_{t},a_{t})$, we denoted the Value function and the Q-value function respectively:
\begin{equation}
    V^{P_0}(s_t)=E\left[\sum_{t\geq0}{\gamma^tR\left(.\middle|s_t\right)|P_o\ }\right]
\end{equation}
\begin{equation}
    Q^{P_0}(s_t,a_t)=E\left[\sum_{t\geq0}{\gamma^tR\left(.\middle|s_t,a_t\right)|P_o\ }\right]
\end{equation}
To determine the optimal set of sub-band frequencies for each hyper-filtering layer, we employed Q-learning algorithms~\cite{sutton2018reinforcement}. By utilizing this reinforcement learning technique, we iteratively trained our system to select the most suitable frequencies based on the observed states and rewards. The results of the RL algorithm, which specify the frequency sets for the two hyper-filtering layers, are summarized in Table II and III. Formally, let be ${W}_{{PPG}}^{i}\left({t}_{k}\right)$ the single segmented waveform of each hyper-filtered PPG time-series (i.e. by using a specific frequency values both in low-pass and high-pass). To capture the dynamic characteristics of the hyper-filtered PPG signals, we computed signal patterns for each sample $s(t_k)$ in the waveform. These signal patterns were derived from analyzing the variations in the hyper-filtered PPG time-series. By examining the changes in signal samples, we obtained a large dataset of signal patterns corresponding to each hyper-filtered PPG signal. The size of this dataset matched the number of filtering frequencies, which was 11 as indicated in Table II and III. To detect the car driver's drowsiness, we extracted a substantial amount of signal patterns from the hyper-filtered PPG time-series. These signal patterns were collected over a time window of 4 seconds, allowing us to capture relevant temporal variations in the PPG signal. The extracted signal patterns served as input to our designed Deep Learning block. In Fig.~\ref{fig5}, we provide a visualization of the generated signal patterns obtained from the temporal variations of the samples $s(t_k)$ for each hyper-filtered signal. These signal patterns effectively capture the unique characteristics of the PPG signal and will be utilized to characterize the driver's level of attention through the subsequent Deep Learning block.

\begin{figure*}
\centerline{\includegraphics[ width=0.8\linewidth]{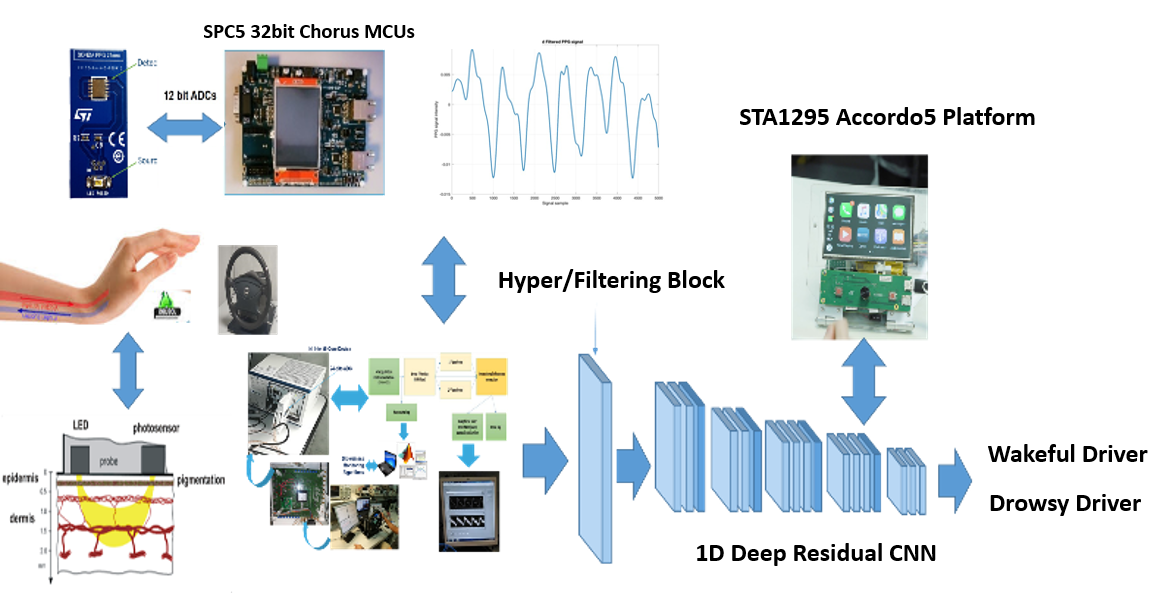}}
\caption{The proposed driver drowsiness monitoring pipeline.}
\label{fig3}
\end{figure*}

\subsection{The Deep Learning block}
As mentioned earlier, we developed a customized Deep 1D Temporal Dilated Convolutional Neural Network (1D-CNN)\cite{bai2018empirical} specifically tailored for our task. This network is designed to process the signal patterns $s(t_k)$ derived from each hyper-filtered PPG signal. The overall architecture of the proposed model can be seen in Fig.\ref{fig3}.
The key innovation of our architecture lies in the utilization of dilated causal convolution layers. The term "causal" indicates that the activation at a given time step depends only on the preceding time step, enhancing the temporal modeling capability of the network. The 1D-CNN comprises a series of residual blocks, with a total of 12 blocks stacked together. Each block consists of a dilated convolution layer, followed by batch normalization, ReLU activation, and spatial dropout. The dilated convolution layer performs a convolution operation with a $3\times3$ kernel. The dilation factor progressively increases for each block, starting from a value of 2.
To complete the pipeline, a two-class softmax layer is added as the final output of the 1D-CNN. This layer predicts the driver's drowsiness level based on the input signal-patterns generated from the hyper-filtered PPG signals.
By leveraging the hierarchical structure of the 1D-CNN and incorporating dilated causal convolutions, our proposed architecture demonstrates the capability to effectively process the hyper-filtered PPG signal-patterns and accurately predict the driver's drowsiness level.

\begin{table}
\caption{Hyper Low-Pass Filtering Setup (in Hz)}
\begin{center}
\resizebox{\columnwidth}{!}{%
\begin{tabular}{|c|c|c|c|c|c|c|c|c|c|c|c|}
\hline
\rule{0pt}{3ex}
\large\textbf{F} & \large\textit{f1} & \large\textit{f2} & \large\textit{f3} & \large\textit{f4}& \large\textit{f5} & \large\textit{f6}& \large\textit{f7} & \large\textit{f8}& \large\textit{f9} & \large\textit{f10}& \large\textit{f11} \\
\hline
\rule{0pt}{3ex}
\large HP & \large0.5 & \large/ & \large/ & \large/ & \large/ & \large/ & \large/ & \large/ & \large/ & \large/ & \large/ \\
\hline
\rule{0pt}{3ex}
\large LP & \large 0 & \large 1.4 & \large 2.9 & \large 2.5 & \large 3.8 & \large 3.9 & \large 4 & \large 4.5 & \large 5 & \large 5.3 & \large 6.9 \\ 
\hline
\end{tabular}}
\label{tab1}
\end{center}
\end{table}

\begin{table}
\caption{Hyper High-Pass Filtering Setup (in Hz)}
\begin{center}
\resizebox{\columnwidth}{!}{%
\begin{tabular}{|c|c|c|c|c|c|c|c|c|c|c|c|}
\hline
\rule{0pt}{3ex}
\large\textbf{F} & \large\textit{f1} & \large\textit{f2} & \large\textit{f3} & \large\textit{f4}& \large\textit{f5} & \large\textit{f6}& \large\textit{f7} & \large\textit{f8}& \large\textit{f9} & \large\textit{f10}& \large\textit{f11} \\ 
\hline
\rule{0pt}{3ex}
\large HP & \large0.5 & \large1.2 & \large2.6 & \large2.7 & \large3.3 & \large3.5 & \large4 & \large4.4 & \large5 & \large5.7 & \large6.4  \\
\hline
\rule{0pt}{3ex}
\large LP & \large7 & \large/ & \large/ & \large/ & \large/ & \large/ & \large/ & \large/ & \large/ & \large/ & \large/ \\ 
\hline
\end{tabular}}
\label{tab1}
\end{center}
\end{table}

\section{Experimental Results}

\begin{figure}
\centerline{\includegraphics[ width=0.95\columnwidth]{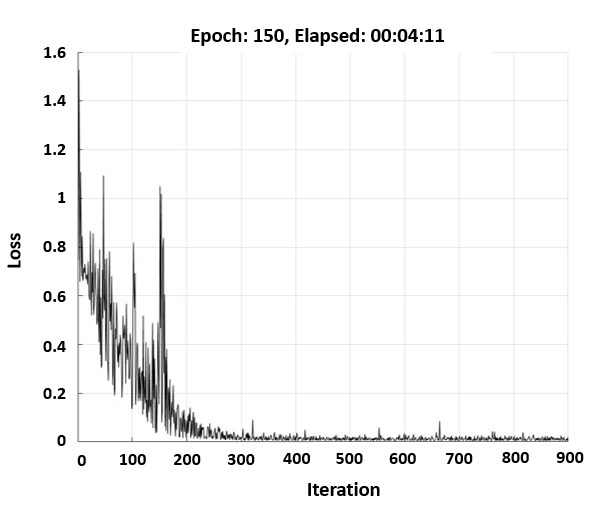}}
\caption{The 1D-CNN Learning Loss dynamic.}
\label{fig4}
\end{figure}

In order to evaluate the performance of our proposed pipeline, we conducted experiments using a dataset consisting of PPG measurements obtained from seventy patients. The age range of the recruited patients was between 21 and 70 years. To validate the effectiveness of our 1D-CNN framework, we collected PPG signals under both drowsy and wakeful conditions. These conditions were supervised and confirmed by experts in physiology, who also acquired the corresponding ECG signals to ensure the subject's level of awareness.
Previous studies have shown that EEG signals can provide insights into the subject's level of attention by analyzing the presence of alpha and beta waves~\cite{rundo2019innovative}. Therefore, we collected PPG signals alongside EEG signals to establish the correlation between physiological signals and drowsiness level. The PPG signals were collected at a sampling frequency of 1 KHz, and the data collection duration was set to 5 minutes.
To perform the experiments, we divided the dataset into a training set (70\% of the total dataset) and a testing/validation set (30\% of the collected data). This division allowed us to assess the robustness and efficiency of our proposed approach. Additionally, we utilized specially designed PPG signals that represented both high and low attention levels to further evaluate the pipeline's discriminative capabilities.
In Table IV, we present the performance of our proposed pipeline compared to other deep learning-based approaches~\cite{rundo2019ad}. The table showcases the effectiveness of our pipeline in accurately classifying different attention levels.
Furthermore, in Fig.~\ref{fig4}, we provide the dynamic learning error of the 1D-CNN, which demonstrates the network's ability to capture the correlation between the hyper-filtered PPG samples and the corresponding attention level of the monitored subjects. This graph illustrates the progressive improvement of the network's performance during the learning process.
Overall, the experimental results validate the robustness and efficiency of our proposed pipeline in accurately identifying and classifying different levels of attention based on the hyper-filtered PPG samples.

\begin{table}
\caption{Benchmark Performance of the Proposed Pipeline}
\begin{center}
\begin{tabular}{|c|c|c|}
\hline
\rule{0pt}{3ex}
\multirow{2}{*}{{\fontsize{8pt}{8pt}\textbf{Table}}}&\multicolumn{2}{|c|}{{\fontsize{8pt}{8pt}\textbf{Car Driver Attention Estimation}}} \\
\cline{2-3}
\rule{0pt}{3ex}
& {\fontsize{8pt}{8pt}\textbf{Drowsy Driver}} & {\fontsize{8pt}{8pt}\textbf{Wakeful Driver}}\\ 
\hline
\rule{0pt}{3ex}
Proposed& $98.71\%$& $99.03\%$   \\
\hline
\rule{0pt}{3ex}
\cite{rundo2019ad} & $96.50\%$ & $98.40\%$   \\
\hline
\end{tabular}
\label{tab1}
\end{center}
\end{table}

\section{Conclusion and Discussion}

Developing a deep learning architecture for classifying driver's attention levels is a highly challenging task, and in our research, we have successfully proposed a 1D-CNN model that outperforms other deep learning approaches in this domain. One major advantage of our method is that it does not require the acquisition of ECG or EEG signals to assess the driver's drowsiness level. This eliminates the issues associated with motion and noise artifacts that can affect the analysis of HRV. Additionally, our approach avoids the need for frequency domain analysis of PPG data, which is required by other HRV-based methods.
We have emphasized the benefits of acquiring PPG signals in a vehicle environment by strategically placing embedded sensors on the steering wheel. The experimental results also demonstrate that our pipeline achieves accurate classification of the driver's attention level with only one minute of PPG acquisition, significantly shorter than the 10-12 minutes typically required by HRV-based methods.
Furthermore, our pipeline can be combined with other promising solutions, such as imaging methods in the visible and infrared spectrum, to further enhance accuracy. The effectiveness of the deep learning framework is evident in its ability to learn signal-patterns generated from the pre-processing of acquired PPG waveforms. Our results clearly demonstrate the effectiveness of the proposed pipeline in assessing the driver's drowsiness level, as evidenced by the reliable performance on the collected dataset of 70 recruited subjects.
Currently, we are in the process of porting the implemented deep learning algorithm to an embedded system based on the STMicroelectronics SoC STA1295 ACCORDO 5, which offers powerful computing capabilities with ARM Cortex cores and a dedicated GPU for graphics and image processing tasks. We are also porting the PPG data filtering and stabilization pipeline to a hardware/software environment based on the SPC5x CHORUS microcontroller technology provided by STMicroelectronics.
In our future work, we plan to explore additional advanced solutions based on deep architectures that have shown promising results in other applications. This includes investigating nonlinear models and oral-based approaches, as mentioned in our previous works~\cite{rundo2018nonlinear, banna2018oral}. By continuously expanding and refining our pipeline, we aim to make significant advancements in assessing and monitoring driver attention levels for enhanced road safety.

\section*{Acknowledgment}
The authors thank the physiologists belonging to the Department of Biomedical and Biotechnological Sciences (BIOMETEC) of the University of Catania, who collaborated in this work in the context of the clinical study Ethical Committee CT1 authorization n.113 / 2018 / PO. This research was funded by the National Funded Program 2014-2020 under grant agreement n. 1733, (ADAS + Project).

\bibliographystyle{IEEEtran.bst}
\bibliography{refs}

\end{document}